# A tool stack for implementing Behaviour-Driven Development in Python Language

Hugo Lopes Tavares[1], Gustavo Guimarães Rezende[1], Vanderson Mota dos Santos[2], Rodrigo Soares Manhães[1], Rogério Atem de Carvalho[1]

[1]Núcleo de Pesquisa em Sistemas de Informação (NSI)
Instituto Federal Fluminense (IFF)
Campos dos Goytacazes, Brasil
`{htavares, grezende, rmanhaes, ratem}@iff.edu.br`
[2]Myfreecomm, Rio de Janeiro, Brasil
`vanderson.mota@myfreecomm.com.br`

**Abstract.** This paper presents a tool stack for the implementation, specification and test of software following the practices of Behavior Driven Development (BDD) in Python language. The usage of this stack highlights the specification and validation of the software's expected behavior, reducing the error rate and improving documentation. Therefore, it is possible to produce code with much less defects at both functional and unit levels, in addition to better serving to stakeholders' expectations.

## 1 Introduction

Software Quality issue has been discussed in a systematic way since early 80s and, since then a series of methods and techniques that aim to ensure the software quality has appeared. More recently, Behaviour Driven Development (BDD) has gained greater acceptance by offering a way which assures that the software functions as expected, it has been adopted by the majority of agile development methods. By origin, BDD aims to integrate verification and validation to a design technic on outside-in style, or better, beginning from the software part perceived by user up to basic units, while it reduces the guaranty costs of software quality, while concentrating on the direct connection of software requisites to the artifact that will implement them: the code.

As BDD strongly bases on the automation of specification tasks and tests, it is necessary to have a proper tooling to support it. Tools are necessary to connect the requirements text to the code, in order to facilitate the testing writing and so on. Thus, this article aims to present a stack of tools created to support the use of BDD in Python, named Pyramid. In order to reach such goal, this paper is divided into a brief BDD introduction, aiming to present its main proposals; followed by the presentation of the tools that compose the stack and, finally, conclusions and future works.

## 2 Behaviour-Driven Development

Behaviour-Driven Development (BDD) is an agile developing technique that encourages collaboration between developers, quality sectors and business personnel in a software project. It was originally conceived by Dan North [4] as a response to TDD [2] limitations, with its focus on the language and interactions used on the process of software development. BDD uses an outside-in approach, or better, the first task is to establish an interaction of software with the user. This way, the acceptance specifications are firstly created then the integration and finally the unit. To develop outside-in implies in thinking, early on how are the clients acceptance criteria after that to think of the design of each part that composes the functionality separately, nevertheless, on the same way that TDD [3], in BDD the tests conduct the software design.

Differently from classical TDD [1], in BDD the unit tests are written in a totally isolated way and for that it is used a technique known as Mock Objects [6] – doubles objects that simulates certain behaviours of real collaborators – while the integration tests are written using real objects, in order to ensure that they interact adequately.

The acceptance and integration specifications in BDD are written using the steps Given, When and Then, described by [4]. *Given* describes an initial context, *When* an event and *Then* an expected outcome – that functions as acceptance criterion. The specifications are named scenarios, grouped in stories or features.

## 3 Pyramid

Pyramid is a stack of tools created to support the employment of BDD in Python language, developed by the Information Systems Research Group, Federal Fluminense Institute (NSI/IFF) since April 2009. Until now, there is no integrated solution that competes with Pyramid, existing only isolated tools for certain BDD tasks, as for example, the Pyccuracy (http://www.pyccuracy.org). This stack has been employed by NSI through the development of systems for Enterprise Content Management (ECM) developed for the Brazilian Federal Government agencies. Pyramid is distributed under the MIT License and is located in http://www.renapi.org/biblioteca-digital/ferramentas, having as main tools PyCukes, Ludíbrio and Should-DSL, as described below.

### 3.1 PyCukes

PyCukes is a command line tool in used to run high level specifications that use *Given, When and Then* of DSL. In PyCukes the specification is written using an external DSL, written in narrative text. The biggest advantage in using narrative text instead of source code is that collaboration between team members may be extended, because any team member can write specifications and pass to developers, who can write implementations required for them. This way, even the client can fix errors and/or make improvements in specifications, so that after the team worry about implementation of the necessary code so that the specification becomes executable. PyCukes supports specifications in Portuguese and English. Code 1 gives an example of specification.

```
Story: Adding content
  As a digital library user
  I want to add content to the library
  So that I can help improve the amount of documents

  Scenario 1: Guest users can't add content
   Given I am at the digital library portal as a guest user
   When I try to add content
 Then I see "Access Denied" error message
```
**Code 1.** Specification in pure text

In order to that specification becomes executable it is necessary to implement the code listed in Code 2 so that step names must coincide with those described in the specification. To run the specifications, just run the tool on the command line.

```
from pyhistorian import Given, When, Then
from should_dsl import should

@Given('I am at the digital library portal as a guest user')
def i_am_at_the_digital_library_as_guest(self):
    browser.open(PORTAL_URL)
@When('I try to add content')
def try_to_add_content(self):
    browser.click('link=Add Content')
@Then('I see "Access Denied" error message')
def ensure_i_see_the_error_message(self):
    "Access Denied" |should| be_into(browser.get_body_text())
```
**Code 2.** Steps implemented in PyCukes

### 3.2 Ludíbrio

Specifying the units of a system is a complex task, because many times the units depend on other components that must not, for any reason, be used in the environment under test. This situation can occur because the components which they depend on are not available, or maybe they will not respond with the expected results upon executing them or because its use would bring on side effects. In this context, Mackinnon, Freeman and Craig [6] define the Test Doubles concept as a technique to support the development, affirming that it encourages the definition of a more structured specification, and improvement of domain code, preserving encapsulation, reducing dependencies and clarifying the interactions between the parties.

Ludíbrio tool is a platform for Test Doubles in Python that provides a simple and expressive way by using an internal DSL for the configuration of Mocks and Stubs [8]. According to Fowler [7], mocks are pre-programmed objects with expectations that form a specification of calls which is expected to be received. When the object under test runs, all expectations defined for the mocks have to necessarily be fulfilled or the specification will fail.

```
>>> def transfer(source_account, destination_account, value):
...         source_account.debit(value)
...         destination_account.credit(value)
>>> from ludibrio import Mock
>>> with Mock() as source_account:
```

```
...        source_account.debit(100) >> None
>>> with Mock() as destination_account:
...        destination_account.credit(100) >> None
>>> transfer(source_account, destination_account, 100)
```
**Code 3**. Example of Mocks with Ludíbrio use

Code 3 shows an example of Ludíbrio use for creating Mocks where it is specified a method *"transfer"* that requires two *"account"* objects. For Ludíbrio use, the declaration *with* allows creating a context, where specification of calls performed on the double is configured and stored with its respective response by using operators ">>" or "<<".

Stubs are objects that behave according to a given logic, as well as mocks. However, unlike these, the execution or not of programmed methods is not a condition so that specifications pass. Normally, stubs are used when the execution of programmed methods is not what the current example wants to specify. The support to stubs in Ludíbrio is done similarly to that shown in Code 3, creating, however, a Stub object instead of a Mock.

### 3.3 Should-DSL

The most basic item of an executable specification is the expectation. An expectation is a statement that is made about an aspect of the software being specified. An example would be to say that, after certain events, *"the source account should have the balance equal to R$ 100,00"* or *"the class should have 20 registered students"*. Expectations can also have a purpose more focused on implementation aspects, such as *"the method math_foo should trigger the exception InvalidNumericalOperation"*.

Traditionally, expectations are implemented using assertion APIs, available in most programming languages in use. Such APIs, however, are too simplistic for heavy usage that makes expectations in BDD. In case of simple expectations, assertions have a good effect, as in Code 4, which shows an example use of standard library Python assertions.

```
assert source_account.balance == 100.00
```
**Code 4**. Python standard assertion

However, assertions do not cover satisfactorily more advanced expectations as *"the plan should contain, among others, the subjects General Chemistry and Linear Algebra"*, as shown in Code 5, using the standard library unittest.

```
expected_subjects = [general_chemistry, linear_ algebra]
for subject in expected_subjects:
    if subject not in grid:
        self.fail()
```
**Code 5**. Non trivial expectation with unittest

Code 5 has an inconvenient: does not have the form of an expectation; cannot be read in a fluent way; fails the specification in a non-natural way, with a call to fail(); it is too much long; its construction is labor-intensive and error-prone; it is not symmetrical [9] in relation to simpler expectations. In addition, following the BDD concepts, it is appropriate that expectations might have, as much as possible, to represent the ubiquitous language [5] of the business being implemented, something that APIs based on assertions usually do not offer. To resolve these issues, an API of expectations called Should-DSL was implemented, which establishes a format for expectations using the term *"should"*. For example, the balance checking code shown in Code 5 would be rewritten by Code 7.

```
Source_account.balance |should| equal_to(100)
```
**Code 6**. Simple Expectation with Should-DSL

```
plan |should| have_all_of(general_chemistry, linear_ algebra)
```
**Code 7**. Nontrivial expectation with Should-DSL

A tool for expectations must also support the implementation of specific matchers for the domain in question. A matcher is the part of expectation that informs what is being scanned. In an expectation *"should be thrown by"*, *"be thrown by"* is the matcher. The Code 8 example shows the creation of a custom matcher for an expectation *"x must be the square root of y"*, which will result in the presented expectation in Code 9.

```
@matcher
def be_the_square_root_of():
    import math
    return (lambda x, y: x == math.sqrt(y),
            "%s is %sthe square root of %s")
```
**Code 8.** Creating custom matcher with Should-DSL

```
3 |should| be_the_square_root_of(9)
```
**Code 9**. Expectation using a custom matcher

## 4. Conclusions and Future Work

This paper presented briefly three tools that compose Pyramid, a stack of tools focused on BDD in Python. These tools are available as Free Software in Python Community and, although they are subject to improvements as the technique and tests demands by automation evolve and other tools are in development to be incorporated to the stack. They are already in daily use in NSI/IFF, as well as by members of Python Development Community, as can be verified by searching the Web.

Regarding the alternatives related to Pyramid tools, in the acceptance specifications arena, there are two alternatives to PyCukes: Freshen and Lettuce. They all use the same external DSL, with minimal differences, and offer like the same features. In Python, there are a number of mocking tools, as PyMock, Mox, Mock.py and others, most inspired on the frameworks Java EasyMock and JMock. Should-DSL has only one alternative turned to BDD, the Hamcrest which is a port of Java homonym tool. The advantages of Ludíbrio and Should-DSL on the alternatives are the same: both have more features than alternatives and both follow a more Pythonic style, or better, more appropriate to languages commonly used in Python language and in dynamic languages in general.

At the current stage, Pyramid have under development a tool called Specloud that customizes the Python unit tests standard library so that you can write unit specifications and obtain results in BDD style. Additionally, a tool that allows the specifications substitution in text for graphical representations of business processes using State Diagrams and UML activities is also under development. Finally, it is important to highlight that soon Pyramid will become the official tooling-reference of the Agile Quality Vector from Brazilian Public Software Portal (http://www.softwarepublico.gov.br/5cqualibr/xowiki/QualidadeAgil).